\documentclass[11pt]{article}
\usepackage{amssymb}
\usepackage{amsmath}

\usepackage{color}
\usepackage{url}
\usepackage{hyperref}
\usepackage{graphicx}
\usepackage{calrsfs}
\usepackage{cancel}

 \newtheorem{rem}{Remark}[section]

\renewcommand{\leq}{\leqslant}
\renewcommand{\geq}{\geqslant}

\begin{document}

\begin{center}
{\bf {\large
VORTICES FOR LAKE EQUATIONS \\(review with questions and speculations) }}
\\ \, \\
Jair Koiller (jairkoiller@gmail.com)\\ \, 
\\
Física Matemática e Computacional, Instituto de F\'isica \\  Universidade do Estado do Rio de Janeiro,     
Brazil
 \end{center}

\noindent MathReviews[2020]: 76B47, 76M60, 34C23, 37E35 \, \\ \, \\
Submitted to Theoretical and Applied Mechanics, Serbia. \\ \, \\
 The `lake equation'  on a planar domain  $D$  with bathymetry  $b(x,y)$  is given by
$   \partial_t u +  (u \cdot  {\rm grad}) u=  -{\rm grad}\, p \,, \,\,{\rm div} (b  u) = 0  \,,\, \text{with}\,\, u \parallel \partial D.$  It  is well posed as a PDE, but  when $b \neq {\rm const}$  justifying  point vortex models requires the analyst's attention.
  We  focus on  Geometric Mechanics aspects, glossing over hard analysis issues.  
Motivating example  is  a  `rip  current' produced by vortex pairs near a beach shore. For uniform slope beach there is a perfect  analogy with  \ Thomson's vortex rings. 
 The  stream function produced by a vortex  is defined as  the Green function of the operator $- {\rm div} ( {\rm grad} \psi/b)$ with Dirichlet boundary conditions.  As   in elasticity,  the  lake equations give rise to  pseudoanalytical functions and quasiconformal mappings.  Uniformly elliptic equations on   Riemann surfaces  could   be   called `planet equations'.


\section{Introduction}  

The   `small' lake equation for a planar vector field $u(x,y)$ on a domain $D$, 
\begin{equation}  \label{eq:lake}
  \partial_t u +  (u \cdot  {\rm grad}) u=  -{\rm grad}\, p \,, \,\,{\rm div} (b  u) = 0  \,\,,\,\,  u \parallel \partial D
  \end{equation} 
  where $b(x,y)$ is the bathymetry,    was proposed  by  Camassa, Holm and Levermore in 1996 \cite{Camassaetal1996,  Camassaetal1997}. It represents a mean field limit  in  shallow waters when
the small vertical component is averaged out. Well posedness as a  PDE was  shown in \cite{Levermoreetal1996}.

For  the modeling  one  pretends  that vertical columns of fluid  move in unison,  so one   adjusts the 2d area form to make  the fluid  
3d-incompressible:  $\rm{div}( b u)  = 0$.  Thus the  velocity fields  $u \in {\rm sDiff}_{\tilde{\mu}}(D)$, with 
$ \tilde{\mu} = b dx \wedge dy . $

  Upon   D. Holm's  suggestion, G. Richardson studied the  lake equations (\cite{Richardson2000}, 2000).   Applying matching asymptotic expansions around a vortex patch of size $ \sim \hspace{-0.3mm} \displaystyle{\epsilon}$  he showed that 
  the patch has   a  self-velocity,  given in  leading order  by  
\begin{equation} \label{eq:Richv}
 \frac{\Gamma}{4\pi}  \log (\frac{1}{\epsilon}) \, \nabla^\perp \log b \,\,\,\,\,\,\,\, ( \nabla^\perp = - i \nabla = (\partial/\partial y , - \partial/\partial x)  \,) .
\end{equation}
For   better approximations, one needs   information about the   inner vortex structure.   
With non constant $b$  a   point  vortex would would move with an  \textit{infinite} self-velocity.  
This contrasts with the    traditional  $b \equiv 1$, where it  is  \textit{finite}.
In that case, the self-velocity is  governed  by the  Robin function $R$  of the  Green function $G_{\Delta}$  for  the Laplacian,  with Dirichlet boundary conditions (see eg. \cite{Crowdy2020}) 
\begin{equation} R(z) = \lim_{\zeta \rightarrow z} \, G_{\Delta}(z,\zeta)+  1/2\pi  \, \log |z-\zeta  |.
\end{equation}

Point vortices for the Euler equations in planar domains  is classic \cite{Lin1941}.   Since  the the 1980's with  Marchioro and Pulvirenti  \cite{MarchioroPulvirenti}    its validity 
 has been consensual. For recent work,  see eg. \cite{SmetsSchaftingen2010, Cao2014}.  
 
 Following the  core energy method  in \cite{FlucherGustafsson1997}, it was proposed  in \cite{KoillerBoatto2015}  that   in closed Riemann surfaces with a metric, desingularization could be done   adding $1/2\pi \, d(p,q)$ to the Green function of the  metric Laplacian.  This  still lacks a real proof  although  it   produces an expected result:   close by vortex pairs mimic a geodesic. Once again, the self motion would be governed by the  Robin function.   
  C. Ragazzo studied  the Robin function of  Bolza's  surface \cite{Ragazzo2017, Ragazzo2022}.   
  
  However, B. Gustafsson \cite{Gustafsson2022} pointed out that when the surface genus is $\geq 1$, the Hamiltonian in \cite{KoillerBoatto2015} is incomplete. One needs also to consider the potential flows and he showed how to describe their  interaction with the   vortices.   
  Together with Gustafsson and Ragazzo  we developed this theme in \cite{Ragazzoetal2024}, which is the basis for this paper. A special case is a multiply connected planar domain, that becomes  a ``pancake'' surface via its Schottky double.

\textit{ The  aim here is to outline how to extend the geometric description in \cite{Ragazzoetal2024}  to the lake equations.}  
  In section \ref{ripsrings} we describe  the  analogy of the classic Thomson's vortex rings with   return (`rip') currents.  
   The  established Hamiltonian description  for the motion of vortex rings, that go back to the 'ancients',  suggests an analogue for the lake equations, presented in section \ref{geomech}.   In  section \ref{greenf} the  stream function of a vortex is introduced: it is  the  Green function of a   non-uniform elliptic  operator. Approximations for the Green function are discussed and a simple toy model for a rip current is presented. Sections \ref{orth} and \ref{red} treat   multiply connected domains\footnote{Our  results essentially coincide with the description in Dekeyser and  Schaftingen
  \cite{DekeyserSchaftingen2020}.}.  Extension to closed Riemann surfaces is outlined in section \ref{planet}, and we finish with two short   comments in section \ref{final}.  
  
We focus only on Geometric Mechanics. Studies about the limits of  validity of  point vortex approximations  were  done for  the lake equation    
  \cite{Valeriola2013, Dekeyser2020, DekeyserSchaftingen2020,  Cao2021, Menard2023, Hientzsch2024}.  They are not 
  discussed here.

 \section{Vortex rings and rip currents} \label{ripsrings}
 
 The infinite self-velocity for  a  `pure'  point vortex  ($\epsilon=0$)   in the lake equation
 is not that  surprising if one is familiar with  William Thomson's (later Lord Kelvin)   torus vortex \cite{Thomson1883, Hicks1883, Dyson1893,Lamb, Hicks1922}.  
 
The   self velocity along the axis  
  is given by
$\Gamma[  \log (8 r/a) - C)/4\pi r $
where $a$ is the inner radius of the torus, and $r$  its radius,  so diverges as  $a \rightarrow 0$.   
The inner vorticity structure determines the value of constant $C$. 

Modern references - also experimental -  are 
 \cite{Auerbach1988, Saf1970,  RobertsDonnelly1970,Leonard1992, LimNickels1992, Nickels1995,Sullivan2008}.  There is a  mathematical proof of existence  (!!)   in \cite{AmbrosettiStruwe1989}. Historical materials can  be found in  \cite{Meleshko2010, Meleshko2011}.  We  call attention to the studies 
  \cite{Gurzhiietal1988, ShashiMarsden2003, Borisovetal2013,  Yang2021}.

What is the connection with lake equations?  In   cylindrical coordinates $(x, r, \phi) $  Euler's equation  for  axisymmetric flows without swirl     is  \textit{exactly the same}  as  the lake equations  on  $(x, y), \, y \geq 0 $ with $b(x,y) = y$, just  making  $r \leftrightarrow y$ and ignoring  $\phi$.    

`Morally'   $a\,\, \text{(from the torus ring)}  \leftrightarrow \epsilon \,\, \text{(the vortex core)}  $ (see \cite{ThorpeCenturioni2000}), 
so there is an  analogy of  rip currents  (informations below) with the   collision of two equal vortex rings,  \cite{LimNickels1992,  ArunColonius2024}. The later is a detailed analysis of the turbulent collision. 
See Fig. 1 and Fig. 2.  

  Imagine two opposite vortices at the same distance from the shore on a slopping  beach, the positive vortex to the right of the negative vortex. 
 As  they move towards the ocean they approach each other. A sequence of such vortex pairs produces the  rip current.   
 The current dissolves at the `head'.  This is   similar (but other  physical mechanism) to the  mutual destruction of  the colliding rings, that produces subsidiary small rings all over. Videos:  \cite{Matsuzawa2023}, \cite{Ryanetal2019}.  
 
As Richard Feynman used to say, ``same equations have same solutions".   Thus  for the lake equations on a uniformly slopping beach one can use \textit{ipsis verbis} the quite extensive literature on coaxial vortex rings.  

\newpage

\begin{figure}[h]
\centering
\includegraphics[width=66mm]{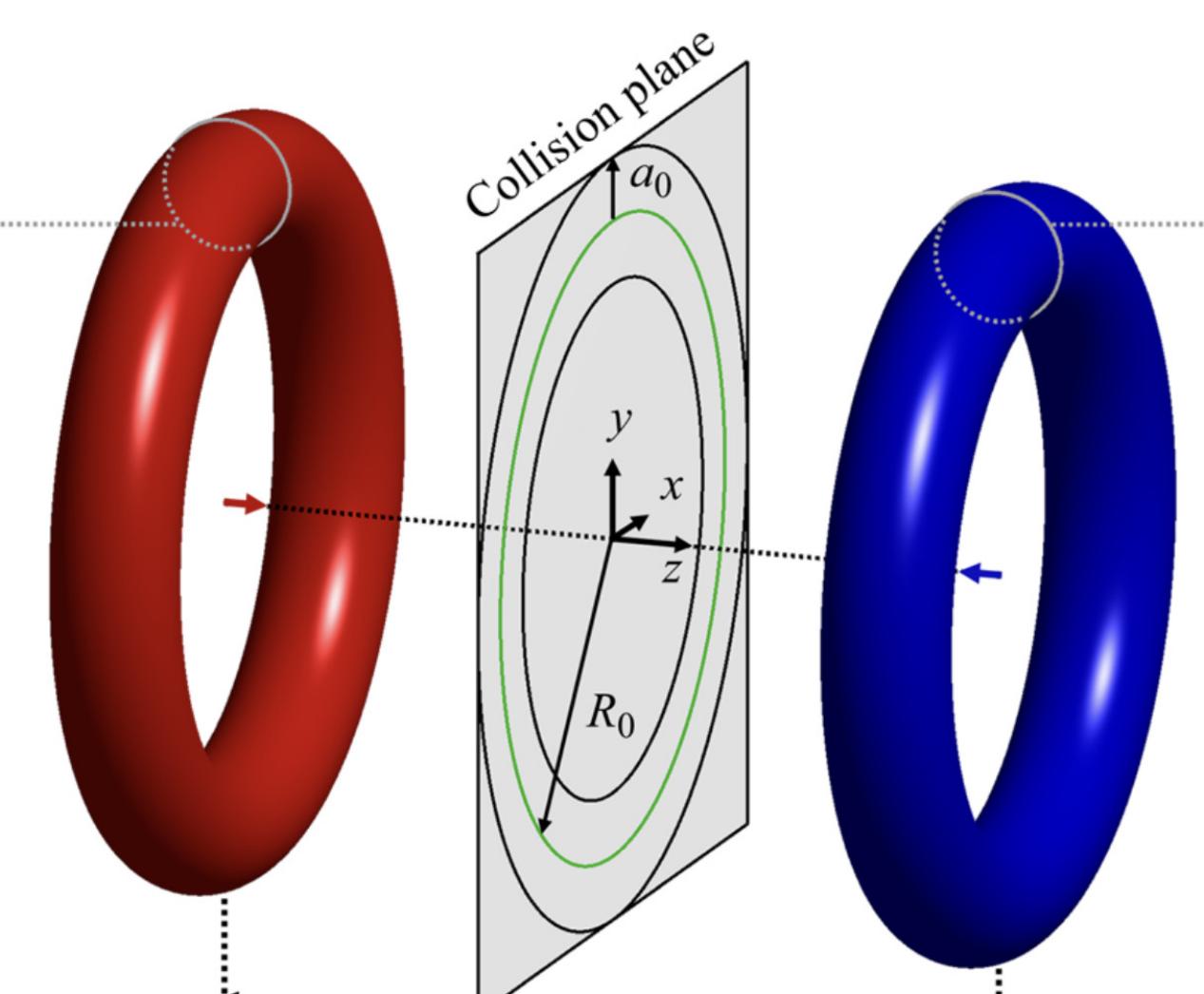}
\caption{Collision of two vortex rings. An important parameter is the thickness. Adapted from \cite{ArunColonius2024}. Recent videos:  \cite{Matsuzawa2023}, \cite{Ryanetal2019}.  
}
\label{fig1}
\end{figure}

\vspace{1cm}

\begin{figure}[h]
\centering
\includegraphics[width=55mm]{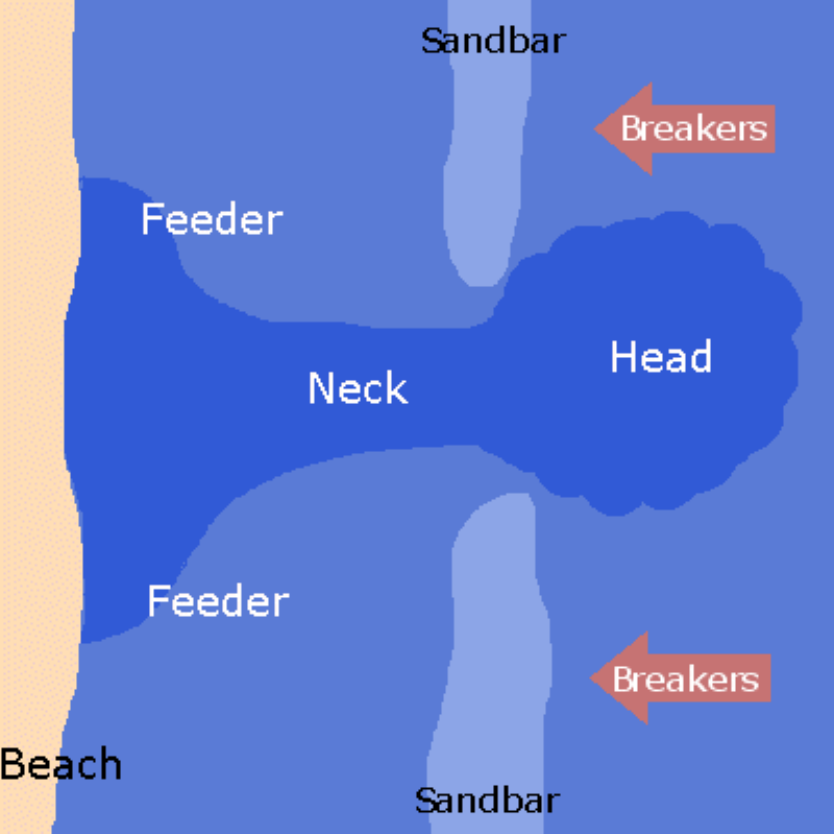}
\caption{Source: wikipedia (public domain).    The wide  head of the current  indicates the
merge of vortex couples.}  \label{fig2}
\end{figure}

\newpage

\section{Geometric Mechanics of the lake equations.\\ Hamiltonian for vortices} \label{geomech}
 
In  Arnold-Khesin  approach  \cite{ArnoldKhesin,Khesin2005}    the energy functional
\begin{equation} 
\frac{1}{2} \int_M  g(u,u) \tilde{\mu} \,\,\,,\,\,\,  u \in {\rm sDiff}_{\tilde{\mu}}
\end{equation} 
is considered on a Riemannian manifold $(M, g)$. The volume form  $\tilde{\mu}$ \textit{can be  unrelated} to the metric $g$ in $M$.   One has   $ {\rm div} _{\tilde \mu} u = 0$  but Euler's  equation  does not change:  
\begin{equation}
u_t +  (v, \nabla) u = -{\rm grad} p   \,\,\,\,\,(\nabla\,\, \text{ is the covariant differential}). 
\end{equation}
Geometrically, it   is  better  treated  dualizing to  ${\rm sDiff}^*_{\tilde{\mu}}$  
 via the   musical isomorphim \textit{relative to $g$},     
  $\, \nu = u^\flat ,  \,  \nu^\sharp = u  $ and Euler equations become
 \begin{equation}
\nu_t + L_u \,  \nu = {\rm exact}\,\,\, \,\,  (u = {\nu^\sharp} \in {\rm sDiff}^*_{\tilde{\mu}} ).
  \end{equation}
The vorticity is defined to be the 2-form 
\begin{equation} \label{eq:vorticityform} 
  \omega = d\nu .
\end{equation}

 Helmholtz's  transport formula 
 \begin{equation} \omega_t  +  L_u  \omega = 0 \,\,\,\, (L_u \, \text{is the Lie derivative})
 \end{equation}
  holds exactly  the same way as in the usual case where  $\tilde{\mu}$ is the volume form of $g$.
  
  \vspace{1mm}
One important consequence is the concept of \textit{isovorticity}, much explored in their treatise:  
\begin{quote}  \textit{``The
vorticity  at any  moment   is transported to the
vorticity at any other moment of time  by a diﬀeomorphism preserving the volume element."}
\end{quote}
\vspace{0.5mm}

In 2d  isovorticity is equivalent to being in  the same  coadjoint orbit \cite{Izosimovetal2016,IzosimovKhesin2017}. 
\vspace{1mm}

 What isovorticity  implies for  a  point vortex model?  
 This is clear:  the ambient \textit{must constant  vorticity}. Usually zero on  a  bounded domain, or a  constant counter-value on  a closed Riemann surface.
 
Therefore  it is natural to define the stream function generated by a  point vortex  to  be the   \textit{Green function} with  that pole. 
\begin{equation} \label{eq:field}
u = (1/b) \nabla^\perp \psi, 
\end{equation}
so that ${\rm div}(b u) = 0$.  The vorticity vector field is 
\begin{equation}  \omega = {\rm curl} (\nabla^\perp \psi/ b) = -  {\rm div} ( {\rm grad} \psi/b)\, \hat{k} . 
\end{equation}
which lead us to  consider  the elliptic  operator, with Dirichlet boundary conditions,   
\begin{equation}  \label{eq:Lb} 
 L_b \psi = -  {\rm div} ( \frac{1}{b} \, {\rm grad}\,  \psi)  .
\end{equation}

The  stream function
of a (bound) point vortex is taken \textit{by definition}  to be the  Green function $ G_{L_b}(z \,; z_o)$  of $L_b$ that we denote $G_b$ for short. 
The  stream function $\psi$  corresponding to  a distributed  vorticity   $\omega$  is recovered  via
\begin{equation} \label{eq:psi}
\psi(z) =  \int_D  \,  G_{L_b}(z,\zeta) \,  \omega(\zeta)   dx dy .
\end{equation} 

 Dualizing to $\nu = u^\flat$  via the Euclian metric,  incompressibility entails
 \begin{equation} \label{eq: incompress}
d( b \star \nu ) = 0  
 \end{equation} 
In the Appendix A of \cite{Ragazzoetal2024} we summarize the properties of   the Hodge star   $\star$,   a conformal invariant object. 
\begin{rem}
We may be going ahead of ourselves, but we recall that when  the domain $D$  is  multiply  connected  the kernel of $L_b$ 
 is non-empty.  
  The  pair of equations  
\begin{equation} \label{eq;pseudoharmonic}
d\eta = 0 \, \text{(being irrotational)}  \,\,\, ,\,\, d( b \star \eta) = 0\,\, \text{(being incompressible)}
\end{equation} defines the  pseudo-harmonic 1-forms. 
For $b \equiv  1 $   
 these are the harmonic form.   
  
 On closed Riemann surfaces those  forms belong to  a $2g$ dimensional space, but
 on  a planar domain only half of those forms on the Schottky double are used, those that are dual to the inner boundaries.
 \end{rem}

 \subsection*{Hamiltonian for vortices on  lake equations. }

The   literature   on   coaxial vortex rings    suggests  a natural proposal.  
\begin{equation} \label{eq:marker}
 u(x,y) = \frac{1}{b(x,y)}  \nabla^\perp \psi \,\,,\,\,\, \psi(x,y ; z_1, \cdots , z_N) = \sum_{k=1}^N \, \Gamma_k G_{L_b}(z, z_k)   
\end{equation}
using  the Green function  $G_{L_b}$  of the elliptic operator  (\ref{eq:Lb}) (more next).   
\vspace{1mm}

The  Hamiltonian system will be 
\begin{equation} \label{eq:hamiltonian}
\begin{split}
& H = \sum_{j<k} \, \Gamma_i \Gamma_k \, G_{L_b}(z_j, z_k) + \frac{1}{2} \sum_j \, \Gamma_j^2  {\rm Rich}_b(z_j)  \\
&  \Omega = \sum_j  \, \Gamma_j  b(z_j) \, dx_j \wedge dy_j.  
\end{split}
\end{equation}
where the  vortex self-velocity  comes from   Richardson's  (phenomenological in  $\epsilon_j$):
\begin{equation}
  {\rm Rich}_b (z_j) = \frac{1}{2\pi} \log (\frac{1}{\epsilon_j}) \, \log b     \,\,\,\,\,\,\,\,  \text{(see \ref{eq:Richv}) } .
\end{equation}

This Hamiltonian system suffices for simply connected domains or for a  closed Riemann surface of genus zero. 
However, the  same way as shown  in   \cite{Gustafsson2022}, it  is  \textit{incomplete}  for multiply connected domains or closed surfaces of  genus $\geq 1$ .  
We outline the ideas for the interplay of vortices  $+$ pseudoharmonic flows  in section \ref{planet}.  It   should  mimic   
 \cite{Ragazzoetal2024}. 
\vspace{1.5 mm}
 
We  end this section with a query: for   consistency   
 it is  important that, when   $b$ approaches a constant value,  then 
 ${\rm Rich}_b(z)$    becomes  the Robin function  
 of  the domain.  What  is correct to do:  to make an interpolation between the expressions or 
to  add the Robin function to  the Hamiltonian (\ref{eq:hamiltonian})?  

\vspace{1.5mm}
  
 \section{Elliptic operators in 2d. \\The Green functions $\boldsymbol{G_b}$ and $\boldsymbol{\Delta_{1/b}}$. } \label{greenf}
  
Elliptic operators  in $\Re^n$   form a  noble area in   mathematical physics   \cite{Garabedian, GilbargTrudinger}.   In divergence form 
 operators with variable coefficients   appear as
\begin{equation}  \label{eq:ellipticA}
L_A \psi = - {\rm div} (A(x) \nabla \psi)  \,\,\,\,, \,\, x \in D \subset \Re^n    
\end{equation}
 where $A(x)$ (\textit{conductivity matrix}) is   positive  and satisfies an uniformity condition.
  In  the now classical  treatises  by  Bergman-Schiffer \cite{BergmanSchiffer} and Vekua  \cite{Vekua1962}  several physics and engineering problems (mainly in elasticity) are taken for motivation.  
 
 For $n=2$, together with Lipman Bers, these authors started the theory of pseudoanalytic functions,   intrinsically connected with quasiconformal mappings. 
 Lipman Bers stands high.  His  1977 review \cite{Bers1977} is  illuminating. 
 There is a review by Henrici \cite{Henrici1957} of Vekua's work.  
 Some references: 
  From  late 40's throughout  50's,   \cite{Bergman1947, Bers1950, Bers1951, Bers1953, Bers1953a, AgmonBers1952, Storvik1957}.   
 From the 60's to early 80's, \cite{Koppelman1961, Rodin1962, Sakai1-1963, Bikchantaev1983}.   A book by Rodin \cite{Rodin1987}  was perhaps the first with the  Riemann surfaces viewpoint. 
From the 1980's to early 2000's interest somewhat subsided, but   now  there is   renewed  interest,  perhaps  motivated by AI and image processing.  Three recent treatises are available:  \cite{Kravchenko2009, Akhalaiaeal2012,Astalaetal2009}.  
Specially exciting   is  the statement  in the latter:    ``any solution becomes
harmonic after a quasiconformal change of coordinates'.  
 
 \section*{Green functions near the diagonal} 
 
 All  operators  act on the first slot. 
  The Green function for  (\ref{eq:ellipticA}) is defined by 
 \begin{equation} \label{eq:GreenA}
\left \{ 
\begin{split}
&  L_A  G_A(x,y) = \delta_y (x)  \,,\,\, x,y  \in D  \subset \Re^n \\ &   G_A(x,y) = 0 \,\,\,  \text{when $y$    is in the boundary}.
\end{split}
\right .
\end{equation}
and $G_A(x,y) = G_A(y,x)$. 
For  lake equations ($n=2$) one  needs   informations about the  Green function (that we denote $G_b$ or $G_{L_b}$)
 for  the  elliptic operator  $L_A$ 
with isotropic conductivity $ A = {\rm diag}\{1/b\} . $

If  the vortices are sufficiently far from   boundaries and the bathymetry changes smoothly,  with  smaller variations compared with their mutual distances, it seems reasonable to use the   dominant term of the Green  function near the diagonal.

For $b \equiv 1$ it is common knowledge that
$G_D(z,\zeta)  \sim  \Phi_0(z-\zeta)$ with
\begin{equation}  \label{eq:phizero}
 \Phi_0(z)=  -\frac{1}{2\pi}\log |z|  . 
\end{equation}

For  the general (\ref{eq:ellipticA})  the authors of  \cite{CaoWan2024b,CaoWan2024} propose the singular + regular decomposition
\begin{equation} \label{CaoWan}
G_A(x, y)=\frac{{\rm det}(A(x)^{-1/2}  + {\rm det}(A(y)^{-1/2} }{2}\Phi_0\left (\frac{T_x+T_y}{2}(x-y)\right )+S_A(x,y),
\end{equation}
where  $  T_x^{-1} (T_x^{-1})^\dagger = A(x). $ They claim that  although the  regular part $S_A(x,y)$ is only  continuous at the diagonal, $S_A(x,x)$ is  $C^\infty$ smooth.

For $ A = {\rm diag}\{1/b\}$ this  becomes
\begin{equation} \label{CaoWan1}
G_b(x, y) \sim \frac{b(x)  + b(y)}{2}\Phi_0\left (\frac{b(x)^{1/2} + b(y)^{1/2}}{2}(x-y)\right ) 
\end{equation}

In \cite{DekeyserSchaftingen2020}  another expression  for the singular part is given   (Proposition 3.1):
\begin{equation} \label{eq:Dekeyser}
G_b(x, y) \sim  \sqrt{b(x) b(y)}\, G_D(x,y) 
\end{equation}
where $G_D$ is the Green function for  the usual Laplacian ($b \equiv 1$) with
 Dirichlet boundary conditions. It is similar to the decomposition in \cite{Donatietal2024}, see Lemma 2.4 and Proposition 2.5 there\footnote{I thank Martin Donati for calling my attention to their paper.}.  \\
 
 These expressions are similar, but not identical. Which one to use?\\
\vspace{-2.5mm}
 
 \section*{Changing the operator   $L_b$ to $\Delta_{1/b}$}
 
It seems  to  us  that the results in  \cite{Khenessy2010}  favor   the choice (\ref{eq:Dekeyser}) more than (\ref{CaoWan1}). The authors
 consider the operator 
\begin{equation} \label{eq:Delta-a}
\Delta_a \psi   =  - \frac{1}{a(x)} {\rm div}( a(x) \nabla \psi ) = - \Delta \psi -  \nabla \log a \cdot \nabla \psi  .
\end{equation}
on a bounded  domain  $D \subset \Re^n, \, n \geq 2$ with   Dirichlet boundary conditions, where   $a(x)$ is  strictly positive and smooth.  $\Delta$ is the Euclidean Laplacian and $\Delta_a$ can be seen as a perturbation of $\Delta$  (look for more below).

By definition, the Green function $\tilde{G} =  \tilde{G}_a$  for $\Delta_a$  satisfies  \begin{equation} \label{eq:GreenDelta}
\left \{ 
\begin{split}
&  \Delta_a  \tilde{G}(x,y) = \delta_y (x)  \,,\,\, x,y  \in D \\ &   \tilde{G}(x,y) = 0 \,\,\,  \text{when $y$    is in the boundary}.
\end{split}
\right .
\end{equation}
and the authors   state  
  that  the symmetry condition  becomes
\begin{equation} \label{eq:symmetrycond}
a(y) \tilde{G}(x,y) = a(x)\tilde{G}(y,x) =: \mathcal{G}(x,y) =  \mathcal{G}(y,x) .
\end{equation}

Let  $\Phi_o$  be  the fundamental solution of the Euclidean Laplacian in $\Re^n$. For $n=2$  is given by (\ref{eq:phizero}). 
They provide a recipe for a recursive  expansion in $x$ for the regular part $H(x,y)$  defined by  
\begin{equation}
H(x,y) = \tilde{G}(x,y) - \Phi_o(x-y)  \,\,\, \text{ (one does not have  symmetry for  $H$)}
\end{equation}

\newpage
Here we do not need  the details of the expansion. $H$   satisfies
\begin{equation}
- \Delta_a H(x, y) = \nabla \log a(x) \cdot  \nabla \Phi_o (x-y)= - \frac{1}{2\pi}  \nabla \log a(x)  \cdot \frac{x-y}{|x-y|^2}
\end{equation}
Although in general  $H   \notin  C^1(D \times D)$, results about elliptic regularity  imply that  is at least $C^o$. Remarkably
the Robin function $H(x,x) $ is  $C^\infty$.
\begin{rem}
 Inverting an elliptic operator smoothes things,  any  solution of a Poisson equation is  two derivatives more regular
than the source  \cite{Real-Oton2022}.  When looking   the literature for lake equations  one must check at start which one was  being considered,  (\ref{eq:Lb}) or (\ref{eq:Delta-a}).
\end{rem}

\vspace{1mm}

In short: for  our purposes it is enough to state that near the diagonal, no matter what anisotropy function $a(x)$ is present,  
\begin{equation}  \label{eq:approx}
\tilde{G}_a(x,y) \sim  \Phi_o(x-y)  . 
\end{equation}

\vspace{1mm}

 We take  $a = 1/b$ in (\ref{eq:Delta-a}). Since   
$    \Delta_{1/b}    = b \, L_b $  with $L_b$ given by (\ref{eq:Lb}) we get the important information that   
\begin{equation} \label{eq:deltab}
L_b \tilde{G}_{1/b}(x,y)  = \frac{\delta_y(x)}{b(x)}  .
\end{equation}
To recover the stream of a vorticity $\omega(x)$ one uses the area form  $dA = b(x) dx dy$:
$$
\psi_{\omega}(z) = \iint_D\, \tilde{G}_{1/b}(z,\zeta)\, \omega(\zeta) dA .
$$
 
 This implies, as  it seems  obvious to us,  that a unit vortex stream function could  be defined alternatively to $G_{L_b}$ of (\ref{eq:Lb}, \ref{eq:psi})   taking
\begin{equation} 
\psi(x) =  b(x_o)  \tilde{G}_{1/b} (x, x_0) . 
\end{equation}  

Let us discuss why.  We are in two dimensions so we compute the circulation of the vector field $\nabla^\perp \psi/b$  around a closed curve $\gamma$ 
bounding a small  region $R$ containing  $x_o$:    
\begin{equation*}
\begin{split}
& \oint_\gamma \, \nabla^\perp \psi/b \,  \cdot d \vec{\ell} = b(x_o) \oint_\gamma \, \nabla^\perp\tilde{G}_{1/b} (x, x_o)/b(x) \, \cdot  d \vec{\ell}  =\\
& \,\,\,\,\,\,\, =  b(x_o) \,  \iint_R      \left(- {\rm div}_x \,  \nabla \tilde{G}_{1/b}  (x, x_o)/b(x) \right) \, dxdy = \\
& \,\,\,\,\,\,\, =  b(x_o) \,  \iint_R    (L_b)_x \tilde{G}_{1/b}(x,x_o)   \, dxdy = \\
& \,\,\,\,\,\,\, = b(x_o) \,  \iint_R  \,  \frac{1}{b(x)} \delta_{x_o} (x)  \,  dxdy  =  b(x_o) \frac{1}{b(x_o)} = 1
\end{split}
\end{equation*}

\vspace{1mm}

This means that  as the vortices $(x_1, \cdots , x_N)$  are allowed to move around the domain by  elements of ${\rm sDiff}$, then  the family of stream functions
\begin{equation}
\psi(x ; x_1 \cdots , x_N) = \sum_{j=1}^N   \, (\Gamma_j  b_j(x_j))   \tilde{G}_{1/b} (x, x_j)
\end{equation}
 is   \textit{ isovorticed with zero background vorticity.} This \textit{ansatz} is close enough to (\ref{eq:Dekeyser}).
 
 \vspace{2mm}
   
In view of (\ref{eq:approx}) we  suggest    the approximation
  \begin{equation} \label{eq:approx1}
  u(x) \sim  \frac{1}{b(x)} \, \sum_{j=1}^N   \,  (\Gamma_j  b_j(x_j))  \frac{ -i (x - x_j)}{\,\, |x-x_j|^2 }   
  \end{equation}
  for the fluid flow produced by    nearby vortices $x_1, \cdots, x_N$. For the vortex dynamics one would take the interaction terms plus each of  Richardson's self velocities.
  \vspace{1mm}
  
 We hope the  above considerations  makes sense.  
An alternative
  Hamiltonian description to (\ref{eq:hamiltonian}), now using  $\tilde{G}_{1/b}$    should be not too hard to produce.  For this, the modified symmetry
  relation (\ref{eq:symmetrycond}) would be instrumental.
  
  \section*{Toy example: vortex pair on a sloping beach}

 In the rigid lid  model it is tacitly  assumed that the momentum of onshore  flow by the surface waves   \cite{BuhlerJacobson2001},  \cite{FokasNachbin2012} is  averaged and  transferred to  alongshore currents,  which  then recirculate via the  rip currents. 
 
 Consider   two opposite vortices on a sloping beach  $b(y) = \alpha y $ initially at $(\pm x_o,y_o)$. The motion will keep symmetry with respect to $y$-axis, so we can focus on the right  vortex  with negative vorticity. The par will   move offshore as they approach.  Then  (\ref{eq:approx1}) gives
  \begin{equation}
  \dot{x} = - p \,  \frac{\Gamma/2}{y}\,\,\,,\,\,\, \dot{y} = \frac{\Gamma/2}{x}
  \end{equation}
  where $p = |\log \epsilon|/2\pi $ is a  phenomenological parameter. We posit  $p = p(\alpha)$ in a decreasing fashion, so that $p \to 0 $ as  $\alpha \to 0$.
  \vspace{1mm}   
  
It is readily seen that in finite time $x(t)$ attains zero while  $y(t) \to \infty$.  Thus rip currents can go very far. 

A nice (and  more sophisticated)  problem  could be  \textit{vortex sheets} on  the lake equations.  Possibly  \cite{Kambe1989} and  \cite{Protas2021} about  Birkhoff-Rott may  help. 
\vspace{2.5mm}

For amusement we  give  information on rip currents.  Not quite an amusement:  
rip currents  cause most deaths on beaches. There is a large  oceanography literature  \cite{WijnbergKroon2002, Dalrympleatal2011, Catelleetal2016, Flochetal2018, Houseretal2020, CastelleMasselink2023} and some modeling by fluid mechanicists with various degrees of sophistication,  \cite{Arthur1962, Peregrine1998, JohnsonMcDonald2004, Hindsetal2007,  Uchiyamaetal2010, McWilliamsetal2018}.  

\vspace{3mm}

\begin{figure}[htb]
\centering
\includegraphics[width=100mm]{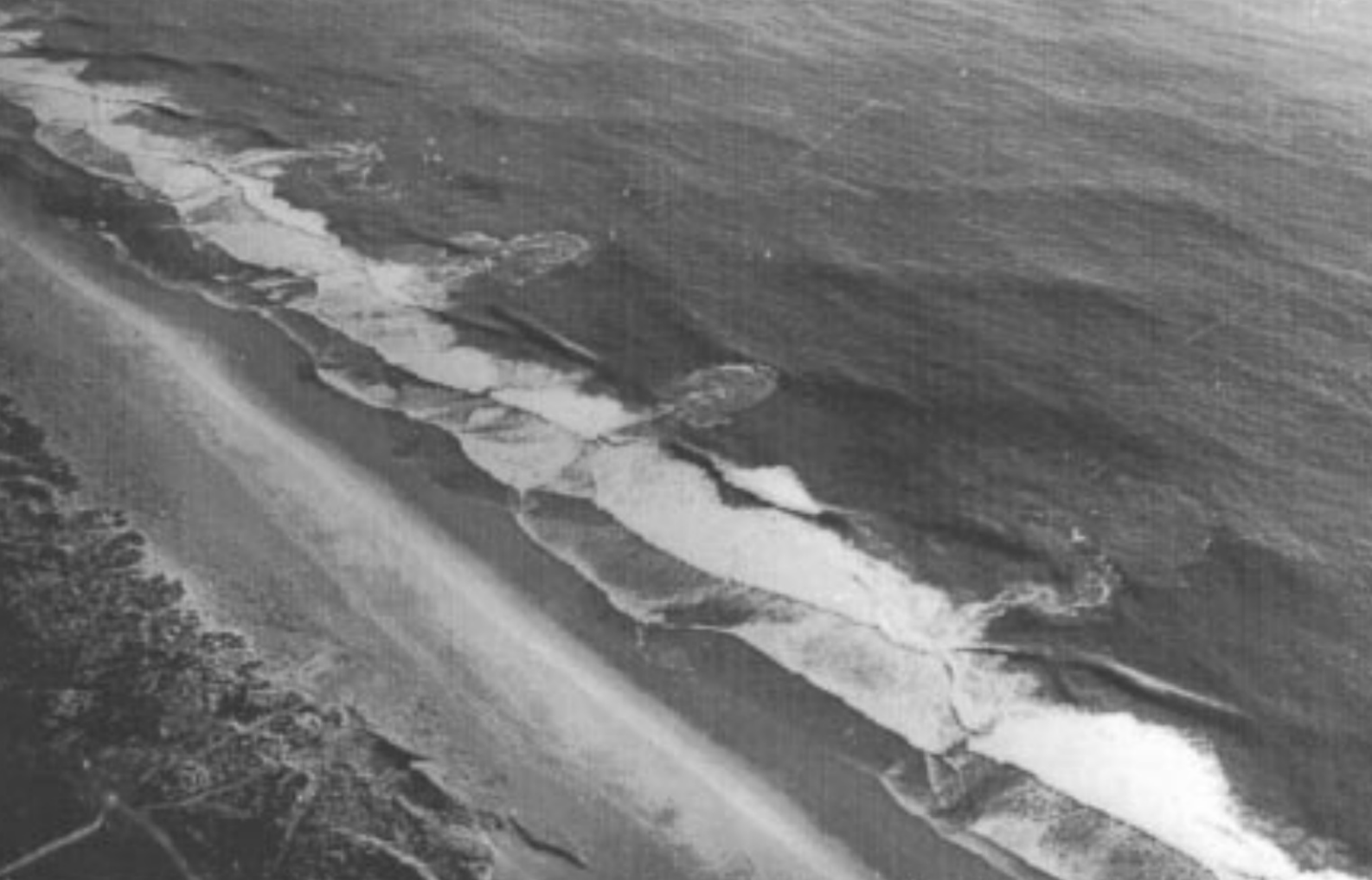}
\caption{Rip currents periodic pattern \cite{Peregrine1998}.   
 Clearly visible are the necks and heads.  The wide   head of the currents indicates the
merge of  eddy couples. }  \label{fig3}
\end{figure}

\begin{quote}
``Peregrine (\cite{Peregrine1998}, 1998) presents evidence for vortex structures arising from
along-shore currents and argues that, in particular, rip currents arise from the
pairing of opposite signed vortices in the form of a propagating dipole. 
(...)  
for typical parameter values, vortices evolve
with length scales of O(100 m) and time scales of O(100 s)"  (quoted from   \cite{JohnsonMcDonald2004}). 
 \end{quote}

\newpage
  
  \section*{Three  Green functions in (more or less)  closed form} 
  
   \emph{Linear profile.}   We discussed in the introduction the  analogy  with vortex  rings. Among the  references we highlighted  the surveys  \cite{Meleshko2010,Meleshko2011} 
and  the  research articles  \cite{Gurzhiietal1988, ShashiMarsden2003, Borisovetal2013,  Yang2021}.  
 From the latter, only changing notation,  
\begin{equation}\label{eq:ring_Ham}
\begin{split}
& \Gamma_j y_j \dot{x_j} =\;\;\frac{\partial H}{\partial y_j}\,\,\,,\,\,\, 
\Gamma_j  y_j \dot{y_j} =-\frac{\partial H}{\partial x_j} \\
& H=\sum\limits_{j=1}^N\frac{\Gamma_j^2}{4\pi} y_i\left[\log\frac{8y_j}{\epsilon_j}-\frac{7}{4} \right]+U(x_1,y_1, \cdots , x_N,y_N) . 
\end{split}
\end{equation} 

For the lake equations one should only replace the self term by
the expression given by Richardson.
The interaction energy  is 
$$ U=\frac {1}{2\pi}\sum_{j\neq i}\Gamma_i\Gamma_j G(x_i,y_i,x_j,y_j)$$ with 
\begin{equation}  \label{eq:green_ring}
\begin{split}
& G(x,y,x',y')=\frac {yy'}{4\pi}\int_0^{2\pi}\frac{\cos\theta\;d\theta}{\sqrt{(x-x')^2+y^2+y'^2-2yy'\cos\theta}}.
\end{split}
\end{equation}
$G$  can be traced back to Dyson 1893 \cite{Dyson1893}. See Lamb's Hydrodynamics \cite{Lamb}. 
\vspace{1mm}

The authors of  \cite{Borisovetal2013} mention the work 
 of Vasilev in 1913 \cite{Vasilev1913}  as the first with the Hamiltonian description.   The integral can be evaluated explicitly in terms of elliptic functions,
 equations (1.2, 1.3), page 36.  The expression they use for the
 self velocity, equation (1.7), page 37, is the one given by Saffman   \cite{Saf1970}. 

\begin{rem}
For axisymmetric generalizations in  higher dimensions,  see \cite{AlexanderWeinstein1953} and \cite{Fryant1981}. For the Geometric Mechanics viewpoint
   \cite{Khesin2013,AntonBoris2018}.
 \end{rem}

\vspace{1mm}

  \emph{Exponential profile.}  These are sometimes used in physical oceanography \cite{Send1989}.  
Inverting the notation of   \cite{GrimshawYi1991}  let $x$ be alongshore ($-\infty < x < \infty$) 
  and $y$  pointing offshore   ($0<y<\infty$). The  profile was given by
  $$
  b(y)  = \left\{  \begin{array}{l}  b_1 \exp[s(y-\ell)], \, 0<y<y_o\\ b_1 \exp[s(y_o-\ell)],\, y> y_o . \end{array}  \right.
  $$
  The Green function $G$ was evaluated  first doing a Fourier transform in $y$. The transformed
  $\hat{G}$ could be found explicitly.
  However, for the   Fourier inversion some approximations were needed.\\
  
\emph{Composite materials conductivity.}  
Suppose  the function $a(x)$ in (\ref{eq:Delta-a}) is formed by a number of piecewise
constant indicator functions in $D$,  
\begin{equation}
a(x) = \sum_{\ell=1}^N \, \kappa_{\ell} \, \chi_{{B}_\ell} + \chi_{{B}_o}\,\,,\,\,\,  \kappa_{\ell} > 0, 
\end{equation}
where the $ {B}_\ell$ are disjoint regions  inside $D$ and
$$
B_o = D - \cup_{j=1}^N \,  B_j
$$

 In \cite{DongLi2019} the problem with two balls is examined.  The Green function is (more or less) explicitly constructed
 (Proposition 2.3) and derivative estimates are provided.

  \section*{Numerical methods for  elliptic PDEs in inhomogeneous media}
  
Numerical analysis of elliptic PDEs is a much developed area, see eg.  \cite{Horvath2013}.  
Here we  just  take from the  lecture notes from a class at MIT    by
S. Johnson, \cite{Johnson2010}.  
He refers to the  Euclidian Laplacian  as  governing
  ``empty space" on a domain $D$, and assumes that its Green function $G_o$    can be constructed.

One  considers  the equation $L_A \psi = f$  with $L_A$ given by (\ref{eq:ellipticA})  in the isotropic situation  $A = {\rm diag}\,  a . $ For us, $a = 1/b$. 
In electrostatics $\sqrt{a}$ is  proportional to the refractive index; in a stretched  drum $a$  is proportional elasticity;  $a$ could be a diffusion coefficient or a thermal conductivity. 

His approach  is to make the problem look as an empty space one, rewriting as
\begin{equation} \label{rewriteDelta}
- \Delta \psi = \frac{f}{a} + \nabla \log a  \cdot \nabla \psi
\end{equation}
Formally
\begin{equation}
\psi(x) = \int_D\, G_o(x,\xi) \left[ \frac{f(\xi)}{a(\xi)} + \nabla' (\log a(\xi)) \cdot \nabla' \psi (\xi)   \right] d{\rm vol}(\xi) 
\end{equation}

This is a volume integral equation for $\psi$. One may rewrite the VIE as 
\begin{equation}
\begin{split}ß
& \psi = \phi_o + B \psi   \,\,\,,\,\,\,
  \phi_o = \int_D\, G_o(x,\xi)   \frac{f(\xi)}{a(\xi)} \, d{\rm vol}(\xi) \\
  &  B \psi = \int_D G_o \nabla' \log a  \cdot \nabla' \psi\,d{\rm vol} . 
\end{split}
\end{equation}

One can think of the inhomogeneous solution as the sum of “homogeneous” solutions using $G_o$ with   right-hand-sides $f(\xi)/a(\xi)$   plus a “scattered” solution due to inhomogeneities of $a$.  
There are well established numerical methods to solve  “VIE” problems.   There are   situations where it  simplifies, for instance:

\subsection*{Piecewise homogeneous media}  This is the third example in the previous section.
Suppose $a \equiv a_1$ in a subdomain $\Omega \subset D$ and $a \equiv a_2$ outside. 
Then  $\nabla a$ is a delta function at the interface, multiplied by  $\log(a_2/a_1)$.  The VIE becomes a SIE which can be handled numerically
more easily. \vspace{1mm}

\emph{This seems promising for the  lake equations,  taking a number of level curves of the bathymetry, and assuming constant values in between them.}
\vspace{1mm}

\subsection*{Born-Dyson approximation}

When the operator $B$  has some norm $< 1$, the functional equation
$ 
(I - B) \psi = \phi_o 
$ 
can be solved via
$$
(I-B)^{-1} = \sum_{k=0}^\infty \, B^k . 
$$
This  is so common in mathematics that there is no name.  In  physics is called  the Born-Dyson expansion (another Dyson).
For nearly homogeneous $a$, one can take 
$$
\psi \sim  \psi_o + B \psi_o \,\,\, \text{(this is another way to justify (\ref{eq:approx})). }  
$$
 \emph{We  wonder:   taking for  $f$ a delta function,  so $\psi_o = G_o$,  perhaps one could produce useful approximations for the Green function of $\Delta_a$.} 

\section{Multiply connected domains: \\orthogonal (Hodge) decomposition} \label{orth}

When the domain has islands, the stream function  (\ref{eq:marker}) is incomplete. In order to    enforce  prescribed   boundary circulations another term is needed,
\begin{equation}  \label{eq:Hodge} 
 \psi(z,t) = \sum_{k=1}^N\, \Gamma_k \, G_b(z, z_k) +  \psi_{\rm circ}(z ;  p_1 , \cdots , p_g) 
\end{equation}

This was called ``outside agency"  by C. C. Lin \cite{Lin1941}, but  it is a  \textit{bona fide} internal entity of the flow. The vortices are  driven  by $ \psi_{\rm circ}$,  but in feedback they change $ \psi_{\rm circ}$ dynamically.  
The process is described in \cite{Ragazzoetal2024}  when  $b\equiv 1$. 
We now emulate  these results for the lake equations.The first step is to show  that (\ref{eq:Hodge}) is  an orthogonal decomposition with respect to the area form $dA= b dx dy$.  
 
\subsection*{First term: $\boldmath{\psi}$ vanishing in all boundaries. }

Let $L_b$ act on $C^{\infty}$ functions with constant (perhaps all different) values on the boundaries.

 Recall  Green's first identity  
\begin{equation*} \label{eq:Green} 
 \int_D \,   {\rm div}( \psi X)  \, dxdy  =  \int_D ( \psi \, {\rm div} X + X \cdot \nabla \psi) dxdy   
=  \,\, \oint_{\partial D} \psi (X \cdot \hat{n}) d\ell   \end{equation*}
Consider  two  functions  $\phi, \psi$ that \textit{vanish on all boundaries (or at infinity).}
 Let  $X_{\phi} = \nabla \phi/ b $.    Interchanging the roles of $\psi$ and $\phi$, it follows that  
$$
\int_D  \psi \, {\rm div} (  \nabla \phi/ b ) dA - \int_D  \phi  \, {\rm div} (  \nabla \psi/ b ) dA = \oint_{\partial D} \cancel{ \psi} (\nabla \phi/ b \cdot \hat{n}) d\ell  -
\oint_{\partial D}  \cancel{\phi}  (\nabla \psi/ b \cdot \hat{n}) d\ell  = 0 .
$$
So  $L_{b}$ is symmetric at least for functions  $\psi, \phi$ that vanish on all boundaries.\\
 
\noindent Now we take only  $\psi$ \textit{vanishing at  all boundaries. }   
Green gives for $X_{\phi} = \nabla \phi/ b $ that  
\begin{equation*}  \label{inner}
 -  \int_D  \psi \, {\rm div} ( \frac{\nabla \phi}{b}) dx dy =  \int_D   \frac{\nabla \phi}{b}   \cdot \nabla \psi    \,  dx dy   =  \int_D   \frac{\nabla \phi}{b}   \cdot \frac{\nabla \psi}{b}  \,\,  b dx dy  
\end{equation*}  
 \centerline{ (in the inner products  
  we can replace $\nabla$ by $\nabla^{\perp}$ and the formulas remain valid).}  
\begin{equation}  \label{inner}
  \int_D  \psi \, \Delta^{b} \phi dx dy =    \int_D  \phi\, \Delta^{b} \psi dx dy  =   \int_D   \frac{\nabla^{\perp}  \phi}{b}   \cdot \frac{\nabla^{\perp}  \psi}{b}  \, b dx dy 
   \end{equation} 
\noindent  This is the  inner product of  $b$-divergence free vector fields, when the stream function of at least one of them (say, $\psi$)
vanishes on all the boundaries $\partial D$. 
 With $ \psi $ vanishing on all boundaries
\begin{equation*} \label{kinetic0}
\langle L_b \psi , \psi   \rangle =:  \, \int_D \,  \psi \,   L_{b}  \psi \, dx dy \, =      \int_D    |\nabla \psi/b|^2 \,  b \,dx dy
\end{equation*}
 which is  twice  is the  kinetic energy of  the  vector field $  \nabla^{\perp}  \psi/b$.  Notice the presence of the area form  $dA = b dx dy $.
 
\begin{rem}  Any $\psi$ vanishing on all boundaries can be described with  the Dirichlet Green function $G_b$ from  its vorticity $\omega$, (\ref{eq:psi}).  The first term in (\ref{eq:Hodge})
corresponds to  this situation, but with concentrated vorticities.  Lacking a better name we call these functions  \textit{pure vorticity} type.  
\end{rem}

\subsection*{ Second term: $\boldmath{ \psi_{\rm circ}(z)}$  are   $\boldmath{b}$-harmonic functions}   Let   $D$ be   bounded with    internal curves $\gamma_i, \, 1\leq i  \leq g$  and   external $\gamma_o$.  
The   kernel of $L_b$  has dimension $g$.   This  follows   from   Fredholm theoryl\footnote{
V. Guillemin,  Elliptic operators \url{https://math.mit.edu/~vwg/classnotes-spring05.pdf}.} or  invoking a continuation starting from $b\equiv 1$.

The $b$-harmonic functions  extend to  the complex domain as pseudoanalytic functions.
 For now it is enough to say that a function $\phi $ is $b$-harmonic when both $d\phi$ (being exact) and $\star d\phi/b$ are closed 1-forms.

One can take for basis the \textit{$b$-harmonic measures}  $m_k, 1\leq k \leq g$, the functions  in the kernel of $L_b$  that  value 1 in each one of the inner boundaries and zero on the others.
As in the Euclidean  Laplacian,
$  0 \leq m_{\ell} < 1 $ in the interior of  $ D,\,$ and
$$  \,m_o + m_1 + \cdots + m_g \equiv 1 . $$
  
For  any $\phi \in {\rm Ker}(L_b)$  we call     $X = \frac{1}{\rho} \nabla^{\perp} \phi$ an   \textit{harmonic flow}.  
 $$   \text{Recall: }   
\psi_{\partial D} \equiv 0   \,\,\,\, \Rightarrow  \,\,\,\,  
 \int_D ( \psi \, {\rm div} Y + Y \cdot \nabla \psi) dxdy =  \,\, \oint_{\partial D} \psi (Y \cdot \hat{n}) d\ell = 0 .
$$

Let  $\psi$ vanish in all boundaries and  $Y = \frac{1}{b} \nabla \phi$  \,, \,with  $\,\phi\, $   $b$-harmonic: $ {\rm div}(Y) = 0$.  
$$ 
\,\,\, \text{It follows that}\,\,\, \int_D \,  Y \cdot  \left( \frac{1}{b} \nabla \psi \right) \, b \,dx dy = 0 \,\,\,\, 
\text{(and we  can replace} \,\, \nabla \text{ by} \,\,  \nabla^{\perp} ).
$$  

\textit{We just verified orthogonality between  pure vorticity and $b$-harmonic functions\\ \centerline{(relative to $dA = b dx dy$).}}

\vspace{2mm}

We end this section with the concept of \textit{$b$-capacity matrix} $P^b$ relative to the pseudoharmonic measures.  The $\perp$'s can be omitted. 
\begin{equation} \label{eq:capacity}
P^b_{\kappa \ell} = \int_{\Sigma}  \frac{\nabla^{\perp} m_{\kappa}}{b} \cdot \frac{{\nabla}^{\perp} m_{\ell}}{b} \,  b \, dx dy \,\, \,\,\,\, (1\leq \kappa, \ell \leq g)
\end{equation}
 They are the coefficients of the kinetic energy of  the harmonic part,
 \begin{equation}
 \psi_{\rm circ}(z) =  \sum_{\kappa=1}^g \,  C_{\kappa} \, m_{\kappa}(z) \,\,\,,\,\,\, T_{\rm har} = \frac{1}{2} C  P^b \,  C^\dagger \, . 
  \end{equation}

 \section{Reduction}  \label{red}

  The $C_{\kappa} = C_{\kappa}(t)  $ evolve in time coupled  with the $N$ vortices  dynamics $z_j(t)$.  
 Making use of the Helmholtz conservation of  circulations  on the $g$ inner boundaries,  it should be possible to eliminate the  $C_j$ and obtain a dynamics just for the vortices.  One would amend  the  Hamiltonian (\ref{eq:hamiltonian}) with the capacities (\ref{eq:capacity}).  \vspace{0.01mm}
 \begin{equation} \label{eq:hamiltonian1}
\begin{split}
& H = \sum_{j<k} \, \Gamma_i \Gamma_k \, G_{L_b}(z_j, z_k) + \frac{1}{2} \sum_j \, \Gamma_j^2  {\rm Rich}_b(z_j)  + \frac{1}{2} C P^b \,  C^\dagger   \\
&  \Omega = \sum_j  \, \Gamma_j  b(z_j) \, dx_j \wedge dy_j.  
\end{split}
\end{equation}
How to determine the row vector  $C = C(z_1, \cdots, z_N)$? 
The following calculations mimic     \cite{Ragazzoetal2024} for 
$b \equiv 1$ and is congenial to  the results in  \cite{DekeyserSchaftingen2020} (eqs.  (2.2) to (2.7)).\\

\noindent  (i) Consider the pseudoharmonic  closed 1-forms $\beta_i = \star d \phi_i/b$,  where 
\begin{equation}    \label{eq:mphi}
\phi_i = \sum_{j=1}^g  (P^b)^{-1}_{ij} m_j .
\end{equation}
 
 \noindent 
 
 Claim:  $\{\beta_1, \cdots , \beta_g \}$  
 are  \textit{dual} to the inner boundary curves $\{\gamma_1, \cdots , \gamma_g \}$. 
 $$
 \oint_{\gamma_i} \beta_j =  \delta_{ij} . 
 $$
We put the proof on hold for a moment. \\  
 
\noindent  (ii) Let us rewrite the  decomposition pure vorticity + $b$-harmonic  (\ref{eq:Hodge}) as
\begin{equation}  \label{eq:Hodge1} 
 \psi(z,t) = \sum_{k=1}^N\, \Gamma_k \, G_b(z, z_k) +  \sum_{\ell=1}^g  B_\ell \beta_{\ell}
\end{equation}
and compute the circulations.  For a single  unit vortex $z_o$, the circulation around the boundary $\gamma_{\ell}$ is
 \begin{equation}
 \,  p_{\ell} = - \oint_{\gamma_{\ell}}\, \frac{\star d\psi }{b} =   \underbrace{ - \oint _{\gamma_{\ell}}\, \star  dG_b(z, z_o)/b}_{m_{\ell}(z_o)  } +  \,\, B_{\ell}  \,\,\,\, \Rightarrow \,\,\,\,\, 
   B_{\ell} = p_{\ell} -  m_{\ell}(z_o) 
               \end{equation}
For  $b \equiv 1$ the    underbraced  equality  is well known by experts  and explained in \cite{Ragazzoetal2024}. It should be  valid in general:
$$
  - \oint _{\gamma_{\ell}}\, \star  dG_b(z, z_o)/b\,\,  \underset{?}{=} \,\,  m_{\ell}(z_o)  
$$

For $ N $ vortices, each with its  corresponding strength $\Gamma_{\ell}$:
\begin{equation}
     B_{\ell} = 
  p_{\ell} -  \sum_{j=1}^N \,\Gamma_j\, m_{\ell}(z_j)   
\end{equation}
  To finish one relates the vectors $B$ and $C$.  This is elementary linear algebra. Let $m = (m_1, \cdots , m_g)$  and
  $\phi = (\phi_1, \cdots , \phi_g)$  in (\ref{eq:mphi}) be seen as column vectors of functions:   $m = P^b \phi$.
  Since $\psi_{\rm circ} = C m = B \phi$ it follows that $CP=B$, with $C, B$ seen as row vectors.
  Thus
\begin{equation}
  T_{\rm har} = \frac{1}{2} C  P^b \,  C^\dagger  =  \frac{1}{2} B  (P^b)^{-1}  \,  B^\dagger 
\end{equation}

We call $Q^b = (P^b)^{-1}$ the capacity  matrix with respect to the  dual 1-forms $\beta = \{\beta_1, \cdots , \beta_g \}$    to  the curves $\{ \gamma_1, \cdots , \gamma_g \}.  $

In order to verify (i), we  invoke Green's first identity in the third line:
\begin{equation*}  
\begin{split}
\hspace{1cm}  & \oint_{\gamma_j}  \beta_i = \sum_{\ell=1}^g\, P^{-1}_{i \ell}  \oint_{\gamma_j}  \,\frac{1}{b}\, \frac{ \partial m_{\ell}}{\partial n}    \\ & =  \sum_{\ell=1}^g\, P^{-1}_{i \ell}  \oint_{\partial D}  \,\frac{m_j }{b}\, \frac{ \partial m_{\ell}}{\partial n}      \\
& \,\,\,\,\,\,\,\,\,\,  =   \sum_{\ell=1}^g\, P^{-1}_{i \ell}  \int_{D} \, \frac{ \nabla m_j \cdot \nabla m_{\ell}}{b}   dx dy \\
&  \,\,\,\,\,\,\,\,\,\, = \sum_{\ell=1}^g\, P^{-1}_{i \ell}  \int_{D} \, \frac{ \nabla m_j}{b}  \cdot \frac{\nabla m_{\ell}}{b}  \, b dx dy   \\
& \,\,\,\,\,\,\,\,\,\, =   \sum_{\ell=1}^g\, P^{-1}_{i \ell}  P_{\ell j} = \delta_{ij} \,\,\,    \text{(for cleanless we removed the superscript $b$ in $ P^b$)}
\end{split}
\end{equation*}
 
\vspace{1mm}

\section{``Planet equations" and pseudoharmonic  forms on Riemann surfaces} \label{planet}

In \cite{Ragazzoetal2024}  the study of point vortices  
 on multiply connected planar domains  was included in the general setting of  Riemann surfaces. If the domain has $g$ internal boundaries and one external,  one takes  the mirror image  to form  the Schottky double,
 a closed Riemann surface of genus $g$. 
 
In so doing, the  classical results of C. C. Lin \cite{Lin1941} were  reinterpreted in terms of Hamiltonian reduction,  \`a la Marsden and Weinstein. The harmonic part can be incorporated to the Dirichlet Green function, yielding the \textit{hydrodynamic} Green function  as in \cite{FlucherGustafsson1997}.

We now  introduce a proposal to extend the lake equations to a closed Riemann surface $\Sigma$ with a metric in its conformal class, given a bathymetry $b$. Since the letter $g$ will be used to denote the underlying  metric, the  surface genus
 will be denoted $\kappa$.  We plan to develop this program in a future study. This project may look  outrageous for a  true fluid-dynamicist,  since  many factors  are disregarded.  For instance, we  are making the  rigid lid assumption (no surface waves) and taking  constant density
(no gravity effects and inhomogeneities).  But this was already the case in the lake equations.  

But  there is  also  a mathematical neglect,  even worse perhaps.   ``Tube" effects \cite{Gray2004} are ignored in the equations that will live in the rigid lid $\Sigma$,   
a  curved manifold bounding  an ambient of one more dimension (the depth).   

At any rate, we believe some mathematical interest could still survive, since the main mathematical object is the the    elliptic operator
$$ L_b \psi = {\rm div} ( \frac{1}{b} {\rm grad}  \psi ) $$
where ${\rm div}$ and ${\rm grad}$ pertain to the metric $g$.  Actually divergence of a vector field  can be defined just with a measure $ \tilde{\mu}:\,\, 
 L_v \tilde{\mu} := [ {\rm div}_{\tilde{\mu}}  v ]\, \tilde{\mu}$ and in dimension 2,
 $$ L_v  \tilde{\mu} = d ( i_v \tilde{\mu}) + i_v \cancel{d \tilde{\mu}} = d (i_v \tilde{\mu})
$$

Let us  denote $\tilde{g} = b g $.  As before,  the area form   $\tilde{\mu} = b \mu$,  where $\mu$ is the area form of $g$  adjusts the bathymetry.  Our vector fields will belong to  ${\rm sDiff}_{\tilde{g}}$.  However, 
the nonlinearity of  Euler equation is  done with  $\nabla_ v v $,    the covariant derivative of $g$.   

Given  any function $\phi$, the vector field  $\frac{1}{b} {\rm grad}^\perp \phi \in {\rm sDiff}_{\tilde{g}}$.   The operation $\perp$ is to rotate -90 degrees in the tangent plane, which is well defined by the complex structure.  However,  recall that  in order to define the `curl' of a vectorfield $v$,  it is necessary to take its musical $\nu = v^\flat $ (always with respect to $g$) and define the vorticity as the two form $\omega = d\nu$.
 
\subsection*{Pure vorticity flows}

The Green function $G = G_b$ for $L_b$ satisfies   
\begin{equation}
 \L_b  G_b(s,r) \mu(s) =:  - d_s \left( \star  \frac{d_s G_b(s, r)}{b} \right) = \left( \delta_{r}(s) - \frac{1}{V}  \right) \, \mu(s)  \,\,\,\, ,\, \,   V = \int_{\Sigma} \mu . 
\end{equation}
($\star$ is the Hodge star) and a ``pure vorticity"  stream function is  recovered with  
\begin{equation} \begin{split}
&
 \psi_{\omega}(s) = \int_{\Sigma}\,  G_b(s,r)  \omega(r) \mu(r)  \,\,\,\,\,\,\,  \text{(note that we use the area form of $g$)}  \\
&
 {  L_b  \psi_{\omega}(s) = \omega(s) - \bar{\omega},\,\,\,\,  \bar{\omega} = \frac{1}{V}\,\int_{\Sigma} \, \omega(s) \mu(s) } 
\end{split} \end{equation}
``Pure vorticity"  (PV) vector fields  are  constructed with   stream functions
\begin{equation} 
 v_{\omega}(s) = \frac{1}{b(s)} {\rm grad}^\perp_g \, \psi_{\omega}(s)  \in PV \subset {\rm sDiff}_{\tilde{\mu}}.  
\end{equation}
 and
\begin{equation}
 d \left(  - \star \frac{d\psi}{b}    \right) =  - {\rm div}_g\left( {\rm grad_g \psi}/b \right)  \mu_g   =  - \Delta^\rho \psi  \, \mu_g .
\end{equation}
But they are not enough to represent all elements of ${\rm sDiff}_{\tilde{\mu}}. $

\subsection*{Pseudoharmonic 1-forms and pseudopotential flows}

The operator $L_b$ in functions can be extended to act on differential forms, as in  Hodge theory for the ordinary Laplacian $\Delta = - {\rm div}{\rm grad}$.
Its kernel has dimension $2\kappa$.  Such forms  $ \eta \in {\rm ker} L_b $ are characterized by  
\begin{equation}
d\eta = d (\star \eta/b) = 0 .
\end{equation}
As already  pointed out by Lipman Bers in the 1950's, these conditions are preserved when $\Sigma$ is conformally changed. 

To  any  canonical  homology basis corresponds a dual cohomology 
 basis $\{\alpha  \, \beta \} $   of $2\kappa$ pseudoharmonic forms.  
 They correspond to the \textit{pseudopotential flows} $$
\frac{1}{b} (\star \alpha)^\sharp \,\,\,,\,\,\frac{1}{b} (\star \beta)^\sharp
$$
that interact dynamically with the pure vorticity flows. 
\vspace{1mm}

\textit{ If the the vorticities are concentrated, 
  we speculate that this interplay should be described essentially  in the same way as  \cite{Ragazzoetal2024}.
}

\vspace{1mm}

 Here we just present one preliminary result in this direction.

\section*{Orthogonality}  

We claim that the  $L_2^{b}$-orthogonal  complement with respect to  the area form $\tilde{\mu}$ inside  $ {\rm sDiff}_{\tilde{g} }$  of the pure vorticity vector fields is given (could it be otherwise?) precisely by the pseudopotential flows.  The proof is a simple abstract nonsense. Flats  and sharps will be always for  the $g$-metric; we omit the subscripts. \vspace{1mm}

Let us try to characterize the vector fields $v \in {\rm sDiff}_{\tilde{\mu}}$ such that
$$ 
\int_{\Sigma} \langle  \frac{1}{\cancel{b}}  {\rm grad}^\perp_g \, \psi_{\omega}(s) \,, \  v  \rangle_g \, (\cancel{b}   \mu) = 0 \,\,\,,\,\,\, \forall \,\, \omega \in C^{\infty}(\Sigma) . 
$$
 This is the same as
 $$  
\int_{\Sigma} \langle  {\rm grad}_g \,  \psi_{\omega}(s) \,, \  v^{\perp}  \rangle_g \, \mu =  \int_{\Sigma} \, d \psi_{\omega}(v^{\perp})  \, \mu =  0 
$$
We need this  Lemma: 
$
 d \psi_{\omega}(v^{\perp})  \, \mu  =  d\psi_{\omega} \wedge v^\flat
 $ . \\
\noindent [ Proof:   $g = a(x,y) (dx^2 + dy^2).$  If   $v = v_1 \partial_x + v_2 \partial y$ then  $v^\flat =  a (v_1 dx + v_2 dy) . $ ]\\
 
Thus the condition is
$  \int_{\Sigma} \, d \psi_{\omega} \wedge v^\flat = 0 \,\, ,\,\, \forall \omega \in C^{\infty}(\Sigma). $   Now, 
$$  d \psi_{\omega} \wedge v^\flat  = d (\psi_{\omega} v^\flat  ) - \psi_{\omega} d v^\flat  $$
Since  $\Sigma$ does not have boundary, the condition rewrites as  
$$ 
\int_{\Sigma} \, \psi_{\omega} d v^\flat = 0 \,\,\,\,\, , \,\,\, \forall \,\, \omega \in C^{\infty}(\Sigma) \,\,\, \Rightarrow \,\,\,   d v^\flat = 0 \,\,\, (and \eta = v^\flat)  .
$$
\vspace{1mm}

 We need also  to impose  also  the $ \tilde{g}$ incompressibility  of  $ v = \eta^\sharp$ .\\

We  use $   i_v \mu  = \star  v^\flat  $     (see Appendix A of \cite{Ragazzoetal2024},  eq. (A.4) ). Then for $ v = \eta^\sharp $: 
 $$  
0 = L_v \tilde{\mu} \,\,\, \Rightarrow \,\,\,  d i_v (b \mu) = d (b  i_v \mu ) =  d (b \star v^\flat) =   d ( \star  b \eta) = 0  
$$

Summarizing, the decomposition is of the form
\begin{equation}  \,\,\,   \nu = - \star  \frac{d\psi_{\omega}}{b}\,  \oplus \,  \eta \,\,\,\,\, \text{with}\,\,\,  d\eta= d(\star b  \eta) = 0   .
\end{equation}

Alternatively, introduce 
$$  
 \tilde{\eta} = \star b \eta \,\,\,\, \text{so that} \,\, \,\, \eta = - \star {\tilde{\eta}}/b  $$ 

The orthogonal decomposition rewrites as   
\begin{equation}
 \nu = [ - \star  \frac{d\psi_{\omega} }{b} ]\,  \oplus \,   [- \star  \frac{ \tilde{\eta }}{b} \, ]   \, \,\,\,\,\,  \text{with}\,\,\, d \tilde{\eta} = d ( \star{\tilde{\eta}}/b ) = 0 .
\end{equation}
and
$    d\nu = \omega \,\,\, \text{since the pseudoharmonic part drops out}. 
$ 
\smallskip \smallskip

In order:    to emulate the results in \cite{Ragazzoetal2024}.   One  needs to extend the Riemann relations for a  canonical homology basis, now using  the dual basis  of  pseudoharmonic forms, and  then compute the circulations of the flow on the homology generators.   With this in hand, it should be a royal road to get the coupled dynamics   between the vortices and the  pseudopotential flows.

\section{Final comments} \label{final}

We hope  that extension of \cite{Ragazzoetal2024} to the lake equations, and its generalization to the `planet' equations  as just outlined will proceed  uneventfully. If we succeed we will submit a short note elsewhere.

We also wanted to point out the direct connection of lake equations to the theory of pseudoanalytical functions and quasiconformal mappings. Perhaps a good way to finish  is by quoting Lipman Bers  \cite{Bers1953a}:

\begin{quote}
``Riemann surfaces were introduced by Riemann as a tool in the investigation
of multiple-valued analytic functions. The ideas and methods
of Riemann's function theory can also be used in studying multiple-valued
solutions of linear partial differential equations of elliptic type." 
\end{quote}

\section*{Acknowledgments} {Thanks to the editors of TAM for the invitation to contribute to the jubilee volume. To  Darryl Holm for suggesting a revisit to the  lake equations via the tools of  Geometric Mechanics. 
Conversations with Boris Khesin, Alejandro Cabrera,  Carlos Tomei and Boyan Sirakov were very helpful. Bj\"orn Gustafsson and Clodoaldo Ragazzo should have been coauthors.
I was partially supported by a FAPERJ fellowship to be  a visiting senior researcher at the Physics Institute of the State University of Rio de Janeiro.}

\bibliographystyle{amsplain}

\end{document}